# Implementation of *Self-Organizing Network* (SON) on Cellular Technology based on *Big Data Network Analytics*

Muhammad Firdaus, Raditya Muhammad, and Rifqy Hakimi
*School of Electrical Engineering and Informatics*
Institut Teknologi Bandung
Bandung, Indonesia.
e-mail: radityamuhammad@students.itb.ac.id, muhammadfirdaus@students.itb.ac.id, rifqyhakimi@stei.itb.ac.id

*Abstrak*— The development of cellular technology will be directly proportional to the increasing requirement in various aspects, such as the speed of data transmission (velocity), data variations (variety), and data storage media (volume). The increase in various aspects will have a direct impact on the growth of internet users worldwide. In relation to that, the demand for increased capacity and coverage becomes a necessity because users access the networks using the same resources by utilizing resource sharing mechanism. Self Organizing Network (SON) is one of the solutions to make the system more efficient with guaranteed Quality of Experiment (QoE). SON can create an automated network with a self-optimization, self-configuration and self-healing mechanism. Big Data is leveraged in the network analytics as a reference for decision-making activities in the network automation process. In this paper, SON using Big Data analytics on 5G network is compared with SON on 4G network as the previous network technology. It is suggested that on 5G networks, Big Data changes the SON detection paradigm from reactive to proactive. Detecting problems starts with data collection and analysis of possible network problems. By predicting network problems early, solving network problems can be done faster. Additionally, by using Multi-hop relay, the distance between User Equipment (EU) and Base Station (BS) will be shorter. Therefore, the aim of increasing capacity and coverage can be fulfilled.

*Kata Kunci*— SON, Big Data, 5G

## I. Introduction

Cellular technology is a technology that is developing very rapidly. Cellular technology that exists today is 2G, 3G and 4G. 4G penetration is still growing, especially in Indonesia. 4G is the latest 3GPP standard technology compared to the previous generation. Data speeds in 4G range at 100 Mbps for downlink access and 50 Mbps for uplink access [1]. Data penetration in 4G technology has been very diverse, but the biggest increase occurred in video services (video calls, streaming, etc.).

Now, after 4G technology has become common in the community, the concept of 5G has not yet been standardized. Institutions in the world are now competing with each other to contribute to the development of 5G technology. Until now there has been no standardized definition of 5G, therefore several telecommunications institutions make their respective definitions. According to GSMA Intelligence [2] 5G technology operates at high frequencies, around 6-300 GHz. 5G technology is a combination of several previous technologies (2G, 3G, 4G) which aims to provide a wider range of services in terms of coverage and availability. And increase the spread of the number of cells and devices in the network. Besides that, it has become a habit, that along with the increase in generation it is always accompanied by an increase in the speed of data transfer. In 5G technology, the data rate launched is 1-10 Gbps, with a latency of 1ms. 5G technology has several requirements such as:

- 1-10 Gbps connection to user (end point connection)
- 1 ms latency
- 1000x bandwidth per unit area
- 10-100x number of connection devices
- 99.999% availability
- 100% coverage
- 90% reduction in energy in the network

• The battery has a low power period of up to 10 years

With these specifications, the variety of services running on the 5G network is increasingly diverse. Services that distinguish 5G networks from services running on 4G networks are the Internet of Things (IoT) and Communication Machine-to-Machine (M2M) services [2].

Along with the variety of services that are increasingly diverse, then user penetration will also increase.

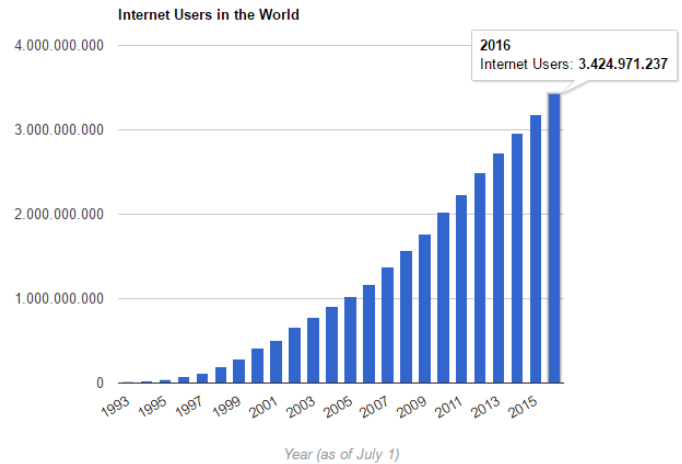

Figure 1. Graph of World's Internet User Penetration [3]

Based on Figure 1, it appears that the penetration of internet users has increased exponentially. The growth of this user must be supported by increasing the capacity and coverage of each cell. Increasing cell capacity and coverage can be done by building a Relay station (RS). RS can increase cell capacity and coverage cheaply and efficiently, later the hospital will be controlled by the Self Organizing Network (SON) which functions to regulate any changes that occur in the hospital [4]

The organization of writing this paper is:
- In part II we explain the supporting theories and classifications of SON based on the literature on existing research
- Part III discusses the function of SON as a solution to overcome the problem of coverage and capacity in cellular networks, as well as its role in cell outage problems in the self healing process.
- Part IV, describes the comparison of the application of SON to 4G and 5G technology, where in the 5G technology Big Data is used as an analytic network.

## II. LITERATURE STUDY

### A. Self-Organizing Network (SON)

SON is an automation technology designed to make planning, configuration, management, optimization and repair of cellular networks to be simpler and faster. SON is expected to be an adaptive intelligent system, where networks can detect changes that occur on the network, and based on these changes SON is able to make decisions to anticipate the change process itself ... From the characteristics of this adaptive system, we can identified that SON has scalability, stability and agility characters. Definition of scalability, to ensure that adaptive algorithms or solutions presented are scalable in achieving minimal network complexity. Stability, ensuring durability when the system is faced with unusual conditions and forcing it to be in an undesirable state, the system must be able to return to the desired conditions in a limited time frame. Agility, is the key character of SON, which not only has the ability to adapt in changing changes, but also responds quickly to that change[5].

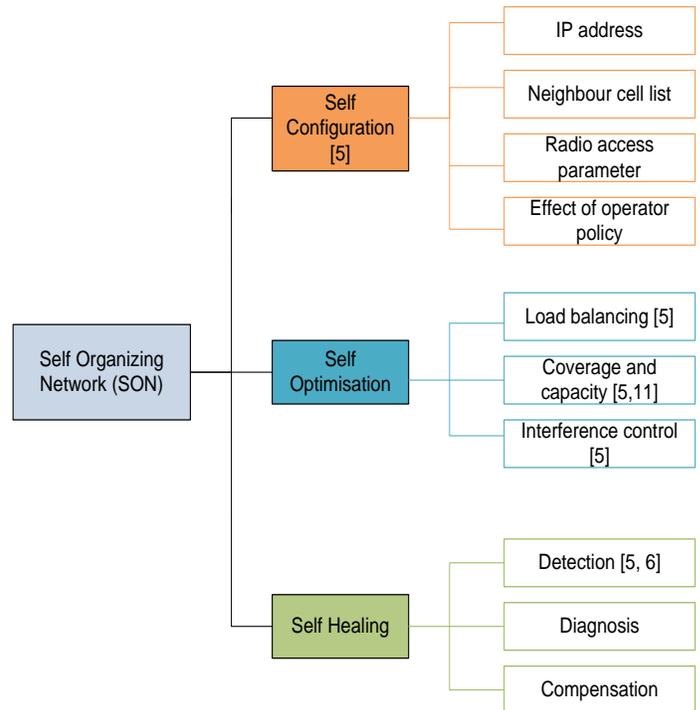

Figure 2. *Self-Organizing Network* (SON) *Taxonomy*

SON is divided into three working mechanisms, namely: self configuration, self-optimization and self healing as described in Figure 2. The mechanism is a system that enables the Telecommunication Network to automatically process, configure, and repair networks. This is where the role of Big Data as a network analyzer that is able to process a number of input data, which is then analyzed based on parameters according to the desired category, so as to produce a set of data as reference material SON perform various actions on a network.

Figure 3 is an illustration that explains how SON works and the relationship between each work mechanism of SON.

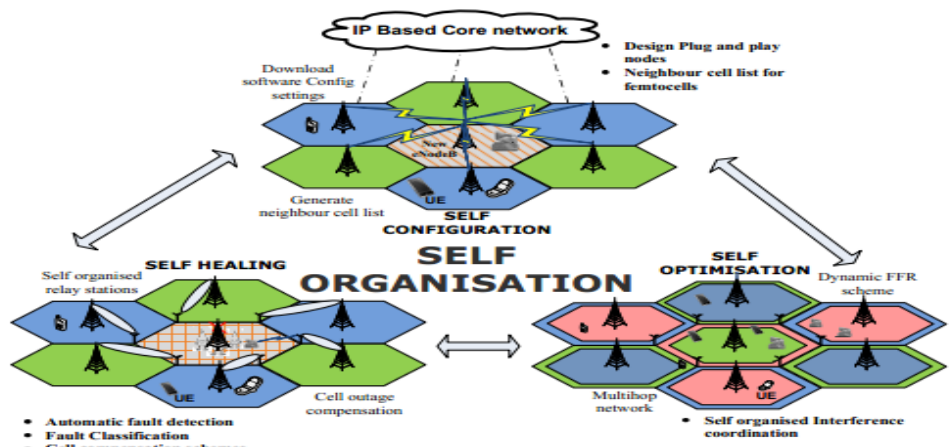

Figure 3. Ilustrasi S*elf Configuration*, S*elf Optimisation* dan S*elf Healing* [5]

*1) Self Configuration*

Self Configuration is a process where the installation process of a new Node is configured automatically with the aim to speed up the process of building a new Node. Configuration is also needed if there are changes in the system, such as a failure in the node, decreased performance, or changes related to the service. eNBs will automatically configure Physical Cell Identify (PCI), transmission frequency and power, making cell planning and rollout faster. So the general framework of self-configuration in future LTE addresses issues related to the construction of new sites without human intervention, but Nodes on the network itself will automatically configure all parameters including IP addresses, neighbor lists and radio access. [5]

*2) Self Optimisation*

After starting with the self configuration stage, it is necessary to proceed to the optimization of system parameters to ensure efficient performance. In cellular systems, optimization is usually done by periodically carrying out drive tests or analyzing the results of log reports generated from the Network Operating System (NOC). But this approach will not be forever maintained in the future network, because the need for a network that can carry out self-optimization will be the main focus of current research. The discussion framework that can be developed in relation to self-optimization includes load balancing, interference control, coverage extension and capacity optimization, to be more clearly seen in Figure 2. However, the focus on this paper is a discussion of self-optimization in increasing coverage and capacity using relaying systems[5].

*3) Self Healing*

In cellular network systems, errors or failures are often found that are caused by non-functioning components or human errors. This failure could be due to software or hardware. Most failures are detected by a centralized Operation and Maintenance (OAM) software and an alarm appears as a warning. When the alarm still appears and cannot be remotely repaired remotely, it must be solved manually by sending a technician who comes directly to the problem site. Of course this requires a lot of time and uncertainty so that the system can return to normal. Even in some cases, there is a failure / error but cannot be detected by OAM software so there will be many complaints from customers. In the Self Organizing Network system, these problems will be overcome by the function of the self healing mechanism. Self healing is a process that can detect, diagnose a mistake remotely, to do recovery actions automatically in an effort to minimize the effects of errors / failures on a network equipment[5].

*B. Big Data* [7]

The Big Data Definition according to ITU-T released in recommendation Y.3600 is a paradigm that allows a broad set of data and has diverse characteristics processed in real time in activities such as collection, storage, management, analysis and data visualization. Meanwhile IBM on its official site defines Big Data in three terms, namely volume, variety, and velocity. The volume here is related to the size of the data storage media that is very large or maybe unlimited. While variety means the types or types of data that can be accommodated are very diverse. While velocity can be interpreted as the speed of the process.

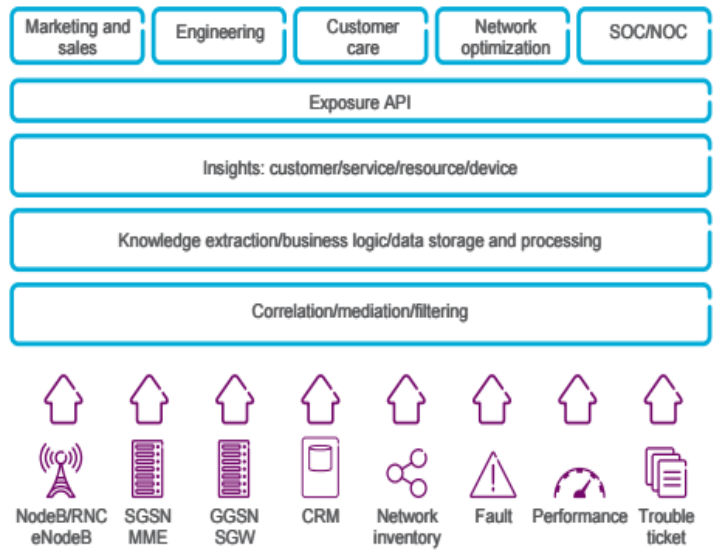

Figure 4. Illustration of Big Data Analytics [7]

SON which has a function as a network optimization, getting the knowledge needed from the results of the analysis of data assets. Data is collected from various data sources. The process of data analysis is done by looking for correlations that are related between data.

### III. SCOPE OF SELF-ORGANIZING NETWORK

In accordance with the previous explanation that the Self Organizing Network has a working mechanism in the form of: Self Configuration, Self Optimization, and Self Healing. But in this paper we focus on the discussion on Self Optimization, and Self Healing. Because in these two mechanisms the level of analysis used by SON plays a very important role. In determining a decision SON takes a number of data from the network which is then processed based on parameters that match the problems that occur, then the results of the analysis become a reference for taking action on a network.

To measure network performance, companies in the telecommunications sector use KPI as an indicator of the quality of the telecommunications network services they have. KPI is in the perspective of telecommunications operators, including capacity, service quality (QoS), capital expenditure (CAPEX), and operational expenditure (OPEX). Then when viewed from the user's point of view, KPI includes service costs, seamless connectivity, infinite capacity and zero latency. The Self Organizing Network is used to bridge the two perspectives to meet the expected KPI [6].

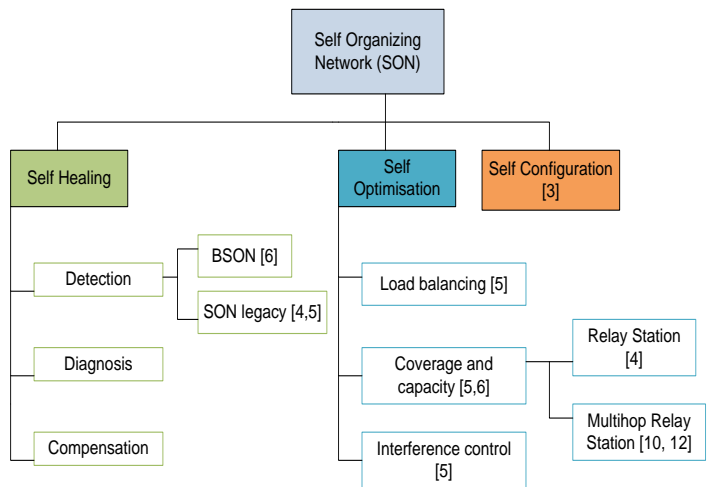

Figure 5. Scope of SON

The Self Configuration described in Figure 2 has components such as: IP address self configuration, Neighbor cell list configuration, Radio Access parameters, self configuration. The process of IP Address self configuration in eNBs is intended so that the new Node can obtain its own IP address, then the central part of OAM will prepare a link to the core network network through the access gateway. This link can be built after checking security and authentication that is on the network core network. After that the Node will download the required software and operational parameters, while the other parameters will be selected automatically based on the previous configuration settings from the neighboring Node. [5]

Neighbor cell list configuration: One of the important functions of Self Organization is the automatic configuration of neighbor cell lists and updating automatically from neighbor relation functions. Each Node has a unique cell ID and an updated list of each neighbor will always be requested when there is a new cell that appears or temporarily checks for a service on a network. [5]

Radio Access parameters self configuration: Configuration of radio access parameters divided into several concentrations including frequency allocation and propagation allocation. Frequency allocation for relays, pico or femtocell, has an effect to minimize interference with existing Nodes automatically. While the propagation allocation is determined by several main factors that influence performance / performance in a network [5]

Self optimization is an important part of the Self Organizing Network (SON). The main goal of self-optimization is to improve performance and efficiency in network maintenance. The framework for self-optimization can be seen in Figure 5. Self optimization is concentrated in three areas, load balancing, coverage and capacity and interference control.

• *Load balancing [5]*

Load balancing is divided into discussion about the resource adaption scheme, traffic shaping, and scheme coverage adaption. In the resource adaption scheme, the main underlying principle is to adjust the amount of resources allocated to the cell to match the traffic load offered. In a traffic shaping scheme, the principle is to form traffic offered to cells so that it can effectively match the traffic load offered with the availability of resources to achieve maximum load balancing. Whereas in the coverage adaption scheme, it is based on a mechanism to change the effective coverage area of the cell to match the traffic offered with the available resources either through the power adapter, antenna adaptation or by both methods.

• *Interference control [5]*

Self optimization provides various ways for interference control and thus can increase capacity. The simplest is to be able to activate certain cells to be deactivated or put in standby mode when not in use. The node then sets the pattern and after a certain period of time in standby mode, the Node will hibernate. This can reduce the interference that affects other cells and also produce energy efficiency.

• *Capacity and coverage via Relay station*

Relaying has been identified by 3GPP as a means of providing effective costs in increasing throughput and coverage optimization. Relay station (RS) is expected to meet two main objectives, namely capacity optimization and expansion of range. According to the paper [12] Relay stations (RS) have two main functions. First, RS can communicate like a customer (user) with a base station (BS) through a backhaul link. On the other hand the RS can communicate like a BS with customers (EU) through an access link. An EU can be connected to a network either via RS that is wirelessly connected to the BS using LTE radio interface technology or directly connected to the BS as in conventional scenarios. As a result, a BS can serve one or several relays in addition to directly serving the mobile terminal. When compared to conventional Base Station (BS), RS is generally cheaper, more energy efficient, simpler and has lower delay. Besides that the main attraction is that repeaters do not have the ability to distinguish real signals from noise and interference. Thus, unwanted signals are also amplified and transmitted along with the original signal.

• *Capacity and coverage via multihop Relay station*

In the past few decades many transmission schemes have been developed to be implemented in cellular networks, some of which are amplify-and-forward (AF) and decode-and-forward (DF). AF is called a wireless repeater or layer 1 (L1), only involves the physical layer. The operation of the hospital can be divided into 2 stages. First, RS receives a signal and then amplifies and forwards it. Operation L1 relay can be on the same carrier frequency (inband) or orthogonal carrier frequency (outband). Another relaying strategy is decode-and-forward (DF), where the signal is encoded by the relay node, then re-encoded and finally forwarded to the desired destination. In this relaying strategy, noise and interference are discarded, but delay will be longer due to the process of decoding and encoding. The structure of relays can be categorized in layer 2 (L2) and layer 3 (L3). Relay station (RS) function can be said only as a device that helps transmit between local BS / eNode B and user equipment (EU). The two main problems in the development of relay technology are the expansion of coverage and improvement in capacity. Multihop links are used to reduce the distance between BS and EU. The use of multihop also helps to achieve higher data speeds than single-hop links. [10]

Self Healing is a SON work process that begins with the process of detecting network conditions, after finding problems with the network, the thing that is done is diagnosing the cause of the error remotely, to perform recovery actions automatically in an effort to minimize the effects of errors / failures caused by human error.

The process of detecting network conditions is a process that greatly determines the length of time repairs carried out by SON. For case studies of sleeping cell problems, that is a condition that occurs in the outer cell which is able to "cheat" SON Legacy because it has poor performance but without generating alarm notifications in Operation And Maintenance Center (OMC). Problems with the existence of poorly performing cells can only be identified when carried out a direct visit to the site, or a manual drive test, or get complaints from customers who feel disadvantaged because they are unable to communicate smoothly [4,5]. SON legacy is reactive: meaning that it can only work to solve problems, when the problem occurs and the impact of the network problem has been felt. This is certainly very detrimental to both companies and users. The process of completing / repairing the network also requires a considerable amount of time, because it is necessary to observe the causes of new tissue failure after a diagnosis of the cause is made, then the solution (compensation) is determined at the final step. The steps that need to be taken create a significant delay so that the use of legacy SON will not be able to meet the 5G network requirements that require as little latency as possible. [5]

From these problems reference [6] proposes a framework for self healing by using Big Data as a component analysis. The problem solving paradigm was changed, from the previous ones who only made repairs when the problem had arisen (Reactive), by using Big Data Self Organizing Network (BSON) the network improvement paradigm was Proactive, meaning that problems that might appear had been detected earlier. BSON uses radio tuners. The process carried out is:

1. Collect data: All data from various sources are combined into a set of data, Big Data.
2. Clasify: determine data based on the level of Operational and Business Objectives.
3. Ranking of KPIs based on KPI values,
4. Dispose of KPIs that are below the threshold value limit.
5. Match the KPI with Network Parameters (NP).
6. Associate the biggest NP when there is a corresponding NP.
7. Vector quantifying with each KPI

By using BSON the detection process can be done faster because mapping of problems that may arise has been done. The process of network diagnosis and repair is done automatically by utilizing machine learning functions. This machine learning works by modeling data, after the data model is formed, optimal network parameters can be identified, which can then be implemented as a network solution.

## IV. METHODOLOGY

In this section we identify possible approaches that have been taken, or can be taken, in analyzing SON based on the SWOT Analysis method. Although Self Organizing Network (SON) has three working mechanisms, in this paper we focus on Self Optimization, and Self Healing. SWOT Analysis (Strength, Weakness, Opportunities, Threats) is a method of analysis that has been widely recognized for auditing and analyzing strategic position analysis objects. With the objectives to be achieved using this method of analysis as a foundation for evaluating potential, and internal limitations, it is also observed the possibility / opportunity of opportunities and threats from the external environment. The process of analysis of Self Organizing Network (SON) is seen from all factors, both positive and negative factors, and in terms of the factors, inside or outside the Self Organizing Network (SON) that can affect the success of implementation. [13]. We use this SWOT analysis to make it easier for readers to know and understand the functions and roles of each branch of the Self Organizing Network (SON). Each branch of SON cannot be directly compared because it has different research objects. So that by using this SWOT analysis, it can be easier to evaluate the advantages, weaknesses, opportunities and threats of each branch of SON.

Table 1. SWOT Analysis of Self-Organizing Network

| SON | Metode | *Strength* | *Weakness* | *Opportunity* | *Threat* |
|---|---|---|---|---|---|
| *Self Optimisation* | *Relay Station* | cheaper, more energy efficient, simpler and has lower delay | The ineffectiveness of the reuse spectrum caused by the additional spectrum needed for the BS-RS access link | Easy to implement on various existing cellular technologies, Potential to be implemented in optimizing smart antennas to overcome the changing mobility of users | The level of efficiency of the frequency spectrum allocated to hospitals cannot be ascertained |
| | *Multi-hop* | cheaper, more energy efficient, simpler and has a lower delay, multihop also helps to achieve higher data speeds than single-hop links. | more complex problems due to backhaul and interaction between users. | Multihop relaying can provide assurance of increased capacity, expansion of coverage area and increased throughput for future technologies, 5G. Prepared to apply the new paradigm to Device to Device (D2D) networking | Fulfilling the requirements of 5G technology, which is 1 ms latency, meets the connection challenges with or without interference from infrastructure |
| *Self Healing* | BSON | Fulfilling the requirements of 5G technology, which is 1 ms latency, meets the connection with or without interference from infrastructure | The addition of infrastructure to support data storage from the subscriber to the core network and other additional data. | Development of the knowledge base that has been obtained to form new business models such as research in the fields of: marketing, government, health, etc. | BSON and Design 5G that are not standardized allow the use of user data that violates privacy. |
| | SON *Legacy* | Compared to the manual self healing process, this SON legacy can shorten the self healing process, much faster. | SON legacy operates to improve the network by finding out the problems that occur on the network first, and then take action. This causes delay. whereas in 5G, SON must be able to predict network conditions. | For 3G and 4G networks that are more tolerant of self healing process delay, the use of SON can reduce OPEX and increase network efficiency. | when running simultaneously on one network, each of the functions of SON allows conflict. whereas in 5G, SON functions increase with the existence of Self-coordination mechanisms that govern each SON |

## V. Conclusions

SON is a network automation technology designed to make planning, configuration, management, optimization and repair of cellular networks to be simpler and faster. SON is expected to be an intelligent system that is adaptive, where the network can detect changes that occur on the network, and based on these changes SON is able to make decisions to take steps to anticipate the changes themselves.

In 5G technology, two things are the main focus: very little latency, and an increase in coverage and capacity. Big Data changed the SON detection paradigm from reactive to proactive. Detect problems starting from data collection and analysis of possible network problems. By predicting network problems early so that solving network problems can be done faster. By using Multi-hop relay shortens the distance between User Equipment (UE) and Base Station (BS), so that the goal of increasing capacity and coverage can be fulfilled.